\def\ba{\begin{eqnarray}}
\def\ea{\end{eqnarray}}

\def\bi{\bibitem}
\def\e{\epsilon}

\documentclass[12pt]{article}
 \begin{document} 
\title{Renyi entropies of a black hole from Hawking radiation}

\author{A.Bialas and W.Czyz\\ M.Smoluchowski Institute of Physics 
\\Jagellonian
University, Cracow\thanks{Address: Reymonta 4, 30-059 Krakow, Poland;
e-mail:bialas@th.if.uj.edu.pl;}\\ Institute of Nuclear Physics PAS, Cracow}
\maketitle

PACS numbers: 04.70.Bw, 04.70.Dy, 05.20.-y

Keywords: Black hole, Hawking radiation, Renyi entropies

\begin{abstract}  Renyi entropies of a black hole are
evaluated by counting the states of the Hawking radiation which fills a
thin shell surrounding the horizon. The width of the shell is determined
from its energy content and the corresponding mass defect. The
Bekenstein-Hawking formula for the entropy of the black hole is
correctly reproduced. \end{abstract}

\section{Introduction}

The effective number of quantum states inside a black hole of mass $M$ is
determined by the  Bekenstein-Hawking formula for its entropy  
\cite{bek,shawk} 
\ba
S_M=4\pi  M^2, \label{bhawk}
\ea
giving an important information about the probabilities $p_i$ of the
states $\psi_i$ forming the black hole\footnote{Throughout this paper we
use $G=\hbar=1$.}. Indeed, when
$p_i$'s are introduced into the general formula for entropy of a
statistical system
\ba
S=-\sum_i p_i \log p_i , \label{ss}
\ea 
they have to satisfy  (\ref{bhawk}). 

To obtain more information about $p_i$ it is necessary to investigate other
quantities of  similar nature. In the present paper we discuss
the moments of $p_i$, i.e.  
the coincidence probabilities,  
\ba
C_l= \sum_{i=1}^\Gamma [p_i]^l
\ea
where $\Gamma$ is the total number of states of the system.

Since the coincidence probabilities are, generally, rather small numbers,
it is more convenient to consider the Renyi entropies \cite{r} defined as 
\ba
H_l=-\frac1{1-l} \log C_l. \label{hl}
\ea
It is  not difficult to show that 
\ba
S=H_1  \label{sh} 
\ea
where the R.H.S. is understood as the limiting value of $H_l$  when 
$l \rightarrow 1$.

Our method of evaluation of $H_l$ is based on the conjecture that the
probability distribution $p_i$ is encoded in the Hawking radiation
emitted by the black hole. To be more precise, we evaluate the Renyi
entropies of Hawking radiation emitted into a thin shell close to the
horizon. The width of the shell is determined from the condition that the
amount of the effective energy emitted into this shell is equal to the
loss of the mass of the black hole.

It is shown that this procedure gives the correct Bekenstein-Hawking
formula (\ref{bhawk}) for $H_1$, i.e. the entropy of the black hole
$[c.f.  (\ref{sh})]$ and  thus it is justified {\it a posteriori}.

Since the straightforward counting of states of the Hawking radiation
leads to a singularity at the black hole
horizon \cite{thooft}, one has to employ a certain regularization
procedure. The singularity comes about because at the horizon ($r=\rho$,
where $\rho=2M$ is the black hole radius and $r$
 is the Schwarzschild coordinate) the $g_{00}$ component of the
metric tensor vanishes:
\ba
g_{00}(r;\rho)=1-\rho/r .\label{gz}
\ea
To deal with  singularities we replace (\ref{gz}) in the region close
to the horizon by 
\ba
g_{00}(\rho\leq r\leq \rho(1+\delta);\rho)=g_{00}(\rho(1+\delta);\rho)
\approx \delta
\label{grho}
\ea
i.e. we take $g_{00}$ to be a constant, starting from a certain
arbitrarily small distance from the horizon. 
For $r\geq
\rho(1+\delta)$ (\ref{gz}) remains valid.
It turns out that our final
result is finite and does not depend on $\delta$ in the 
limit $\delta \rightarrow 0$.
Thus  our regularization procedure does not introduce any additional
uncertainties. 

To summarize, we evaluate the Renyi entropies of the Hawking radiation
in a certain region of configuration space. This region is selected in
such a way that its energy content reproduces the mass lost by the black
hole through the radiation. It turns out that in this way the standard
Bekenstein-Hawking formula for entropy is recovered. The same procedure
is then employed for evaluation of other Renyi entropies. 

In the next section the general formulae for entropies of the Bose gas
are given. They are applied to black holes in Section 3. In Section 4
the appropriate volume is estimated and expressed in terms of mass
defect. The final formulae for the Shannon and Renyi entropies are
obtained in Section 5. Our conclusions are listed in the last section.
Evaluation of some relevant integrals is described in the Appendix.

\section{A general formula for Renyi  entropies of
a Bose gas}

We follow closely the argument given in \cite{bc} where we have obtained
the relation
\ba 
H_l= \left(1+\frac1{l}+\frac1{l^2}+\frac1{l^3}\right) \frac{S}4,\label{hph}
\ea
valid for the free photon gas closed in a large static volume 
(in the flat space). The main result of this paper is that the same
formula is also valid for black holes.

The probability of having $n_1$ bosons in a state with energy $\e_1$, $n_2$
bosons with energy $\e_2$,... is given by
\ba
P(n_1,n_2,...n_M)=\prod_{m=1}^M \left(1-e^{-\beta \e_m}\right)
e^{-\beta n_m \e_m}
\ea
where $\beta=1/T$ and we have put the chemical potential to zero.

The coincidence probabilities are 
\ba
C_l= \sum_{n_1,n_2,...}\left[P(n_1,n_2,...n_M)\right]^l
 =\prod_{m=1}^M \frac{\left(1-e^{-\beta \e_m}\right)^l}
{\left(1-e^{-\beta l \e_m} .\right)}
\ea
This gives for the Renyi entropies
\ba
H_l=\frac1{1-l} \log C_l=
-\sum_{m=1}^M \log\left(1-e^{-\beta \e_m}\right)+ \frac1{1-l}
\sum_{m=1}^M \log\left(\frac{1-e^{-\beta \e_m}}
{1-e^{-\beta l \e_m}}\right).
\ea
Finally, when the sum over states is replaced by an integral we
have
\ba
H_l=\int d N (\e)W_l(\beta \e) \label{hlc}
\ea
where $dN(\e)$ is the number of states with energy between $\e$ and $\e+d\e$
 and
\ba
W_l(z)=-\log(1-e^{-z})
+\frac1{1-l}\log
\left(\frac{1-e^{-z}}{1-e^{-lz}}\right).  \label{wl}
\ea
 This formula  
was  derived in \cite{bc}.

\section{Application to  black holes}

Our problem now is to evaluate $H_l$ for the radiation emitted by the
black hole of radius $\rho$ into an infinitesimal layer around its
horizon. To this end we first have to specify the physical meaning of
the energy $\e$ and of the temperature $T=1/\beta$ in the case of a
black hole. Since the black-body radiation is in equilibrium, 
 the temperature $T$ is a constant, provided $\e$ is the
energy {\it conserved} in the process\footnote{Since the gravitational
field outside of the black hole is static, the conserved energy can be
defined, c.f. \cite{lltp}, Section 88, and \cite{llfs}, Section 27.} 

This  implies that $T$ is the
Hawking temperature 
\ba
T=T_H=\frac{1}{4\pi \rho}= \frac{1}{8\pi  M} . \label{thro}
\ea

To perform the integration in (\ref{wl}) we first have to determine 
the density of states $N(\e)$. The relevant calculation was performed in 
 \cite{thooft} where the formula for the number of states with energy
smaller than $\e$ was given in the form
\ba
N(\e)= \frac{dr}{\pi g_{00}} \int_0^{l_{max}} (2l+1)dl
\sqrt{\e^2-g_{00}\left(m^2+\frac{l(l+1)}{r^2}\right)}
\ea
where $g_{00}$ is given by (\ref{gz}). 
The integration extends for the values of $l$ for which the argument
of the square root is positive. This integral can be evaluated and one
finally obtains  
\ba
dN(\e)=\frac{dN(\e)}{d\e} d\e=
\frac{2 r^2 dr}{\pi [g_{00} ]^2} \sqrt{\e^2-g_{00} m^2}\e d\e
\approx\frac{2 r^2 dr}{\pi [g_{00} ]^2} \e^2 d\e . \label{dx}
\ea

Introducing (\ref{dx})  into (\ref{hlc})   
we obtain for the Renyi entropy 
contained in the layer between $\rho$ and $\rho + d r$
\ba
d H_l =\frac{2 \rho^2 d r}{\pi [g_{00}]^2 }  
\int_0^\infty \e^2 d\e W_l(\beta \e)
=\frac{2 }{\pi \sqrt{g_{00}}}\frac {d r}{\rho}
(\rho T_H)^3 \Phi_l =
\frac1{32\pi^4  [g_{00}]^2}\frac {d r}{\rho} \Phi_l  
\label{hlbiz}
\ea
where $\Phi_l $ are numerical constants defined as
\ba
\Phi_l= 
\int_{\e_0}^\infty z^2 dz W_l(z)\;;\;\;\;    \label{fil}
\ea
with $W_l$  given in (\ref{wl}).

We see that the formulae (\ref{dx}) and (\ref{hlbiz}) exhibit a
singularity at $r=\rho$ and the procedure to regularize this
singularity, described in the Introduction, must be applied. As already
mentioned, it turns out that the final result does not depend on the
details of regularization.

One can also verify that, when formally integrated from $r=\rho +h$ to
$\infty$, the divergent part (proportional to $1/h$) is identical to
that obtained by 't Hooft \cite{thooft}.

\section{Relation to the mass defect}

We have evaluated the contribution to the Renyi entropies from the
Hawking radiation emitted into an infinitesimal layer of width $d r$
outside of a black hole. At this point the width $d r$ is infinitesimal
still arbitrary. To connect it to the physical properties of the black
hole we relate it to the change of the black hole mass, $d M$.

To relate $d r$ to $d M$ we observe that the emission of
radiation causes the decrease of the mass of the black hole by the
amount of emitted energy reduced by the amount of free energy used in the
process of  shrinking of the black hole:

\ba 
d M=d E-d F \label{me}
\ea 
where $F$ is the free energy of the photon gas. 

The amount of emitted energy $d E$ can be evaluated from the
well-known formula for the Bose gas:
\ba
dE=dN(\e)\e\frac{e^{-\beta \e}}
{1-e^{-\beta \e}}. \label{ee}
\ea
Similarly, using the relation between the free energy and the statistical
sum $Z$ we have
\ba
dF=-dN(\e)T\log Z= 
dN(\e)T\log (1-e^{-\beta \e}) .\label{ff}
\ea
Using the formula (\ref{dx}) for $dN(\e)$ we obtain 
\ba
dE=\frac2{\pi [g_{00}]^2}\left(\rho T\right)^3 
\frac {dr}{\rho}T\Omega \label{eee}
 \label{eab}
\ea
\ba
dF=\frac2{\pi  [g_{00}]^2}\left(\rho T\right)^3
\frac {dr}{\rho}T\omega \label{fff}
\ea
where
\ba
\Omega= 
\int_{\e_0}^\infty z^3 dz \frac{e^{-z}}{1-e^{-z}}\;;\;\;\;
\omega=    \int_{\e_0}^\infty z^2 dz \log (1-e^{-z}).
\ea
Introducing (\ref{eee}) and (\ref{fff}) into (\ref{me}) 
we have

\ba
dM = \frac1{32\pi^4  [g_{00}]^2}
\frac{dr}{\rho}T(\Omega-\omega) \label{dmef}
\ea
where $g_{00}=g_{00}(r;\rho)$ is defined by (\ref{gz}) and (\ref{grho}).

Consequently,
\ba
\frac{dr}{\rho} = \frac{32\pi^4  [g_{00}]^2}
{(\Omega-\omega)}\frac{dM}{T_{H}} \label{drhof}.
\ea

This formula gives the width of the shell around the horizon where the
energy content of the Hawking radiation balances the mass defect of the
black hole. To evaluate the Renyi entropies we now use our main idea,
assuming that that not only energy but also entropy lost by the black
hole through radiadion is contained in this shell.

\section{Renyi entropies of the black hole}

When  (\ref{drhof}) is introduced   into (\ref{hlbiz}), 
one sees that  the singular
factors $[g_{00}]^2$ in the numerator and denominator cancel exactly and  
one obtains
\ba
dH_l=\frac{\Phi_l}
{\Omega-\omega}8\pi  M dM \label{dh}
\ea
where we have  used (\ref{thro}). Thus we have expressed the change of
the Renyi entropy in terms of the change in the black hole mass,
essentially repeating the original procedure of Bekenstein
\cite{bek,suss}.

 After integration of (\ref{dh}) from $0$ to $M$ we thus have 
\ba
H_l=\frac{\Phi_l}
{\Omega-\omega}4\pi  M^2 =\frac{\Phi_l}
{\Omega-\omega}S_M,  \label{fin}
\ea
where $S_M$ is the Bekenstein-Hawking entropy of the black hole
\cite{bek,shawk}, given by (\ref{bhawk}).
The numerical constants $\Phi_l, \Omega$ and $\omega$ are
evaluated in the Appendix and are given by  formulae (\ref{phil}),
(\ref{phi1}), (\ref{Om}) and (\ref{om}). It is also demonstrated 
[c.f (\ref{fOo})] that 
\ba
\Phi_1=\Omega-\omega.
\ea 
With this condition (\ref{fin}) implies the correct formula 
(\ref{bhawk}) for the Shannon entropy.
 This verifies  that our conjecture gives the  correct estimate
of  the numer of states inside the black hole.

Using now $\Phi_l$ given in the Appendix we obtain for the Renyi
entropies the formula (\ref{hph}).

\section {Conclusions }

Our conclusions can be summarized as follows.

(i) We have evaluated the Renyi entropies of a Schwarzschild black hole
of radius $\rho$ using the conjecture that they are given by the
entropies of the Hawking radiation which fills a thin shell around the
horizon. The width of the shell is determined from the condition that
the energy of the radiation contained there is equal to the mass defect
which the black hole suffers during the emission process. When applied
to $H_1=S$, this procedure reproduces the correct formula (\ref{bhawk})
for the Bekenstein-Hawking entropy, thus providing the {\it a
posteriori} justification of the argument. The method requires a
regularization procedure for the Schwarzschild metric close to the
horizon, but the effects of regularization disappear from the final
result.

(ii) All Renyi entropies are proportional to the Shannon entropy. The
ratio $H_l/S$ turns out to be independent of the mass of the black hole.
It is interesting that this ratio is identical to that obtained for the
photon gas in a fixed volume.

(iii) Since the Renyi entropies provide additional information about the
statistical properties of the black hole,  our calculation may perhaps
be useful in the search for its  internal structure.

(iv) Our calculation is consistent with the idea that the entropy of the
black hole is determined by the state of its surface
\cite{bek,shawk,suss}. Indeed, it shows that the black hole entropy can
be evaluated from the Hawking radiation at the horizon.

\vspace{0.5cm}

{\bf Acknowledgements}

Discussions with Piotr Bizon, Andrzej Staruszkiewicz and Kacper Zalewski
are highly appreciated. This investigation was partly supported by the
MEiN research grant 1 P03B 045 29 (2005-2008).

\section {Appendix. Evaluation of numerical constants}

To find $\Phi_l$ we have to evaluate integral 
\ba
\Theta_l= \int_0^\infty z^2 dz\log \left(1-e^{-lz}\right) 
\ea
This can be done by expansion in series of $e^{-lz}$:
\ba
\Theta_l= -\sum_{n=1}^\infty\frac1{n}
\int_0^\infty z^2 dze^{-nlz} = -\sum_{n=1}^\infty\frac{2}
{l^3n^4}=-\frac{\pi^4}{45l^3}
\ea
Using this we have 
\ba 
\Phi_l =-\frac{l}{l-1}\Theta_1 +\frac1{l-1}\Theta_l
= \frac{\pi^4}{45} \frac{l^4-1}{l^3(l-1)}
=\left(1+\frac1{l}+\frac1{l^2}+\frac1{l^3}\right) \frac{\Phi_1}4 \label{phil}
\ea
with
\ba
\Phi_1=\frac{4\pi^4}{45}.  \label{phi1}
\ea
In a similar way we obtain
\ba
\Omega=\sum_{n=1}^\infty \int_0^\infty 
z^3 dz e^{-nz}=\frac{\pi^4}{15} \label{Om}
\ea
and 
\ba
\omega=    \int_0^\infty z^2 dz \log (1-e^{-z})=\Theta_1=
-\frac{\pi^4}{45} .\label{om}
\ea
Thus, consequently,
\ba 
 \Omega - \omega =\Phi_1 .\label{fOo}
\ea

\end{document}